\documentclass[11pt,a4paper]{article}

\usepackage[T1]{fontenc}
\usepackage[utf8]{inputenc}
\usepackage{amsmath}
\usepackage{amssymb}
\usepackage{graphicx}
\usepackage{booktabs}
\usepackage{geometry}
\usepackage[super,sort&compress]{natbib}
\usepackage[hidelinks]{hyperref}

\usepackage{siunitx}
\usepackage{xspace}

\newcommand{\Rheadline}{1.207\thinspace507\thinspace039\thinspace343\thinspace337\thinspace716\thinspace0}
\newcommand{\RrawHead}{1.207\thinspace507\thinspace039\thinspace343\thinspace337\thinspace715\thinspace8}
\newcommand{\RsegOne}{1.207\thinspace507\thinspace039\thinspace343\thinspace337\thinspace716\thinspace8}

\newcommand{\RwlsRawT}{1.207507039343337716017}
\newcommand{\RmpRawT}{1.207507039343337716017}
\newcommand{\RbayesRawT}{1.207507039343337716022}
\newcommand{\uStatE}{1.14}
\newcommand{\uStatShort}{1.14}
\newcommand{\uWlsE}{1.128}
\newcommand{\uMpE}{1.128}

\newcommand{\chiRed}{0.421}
\newcommand{\dofV}{15}
\newcommand{\pChi}{\ensuremath{9.7\times10^{-1}}}
\newcommand{\birgeRatio}{0.649}
\newcommand{\xiMpE}{0.00}
\newcommand{\xiBayesE}{0.14}

\newcommand{\dDriftPerDay}{\ensuremath{0.7\times 10^{-20}}}

\newcommand{\dSegNineWls}{\ensuremath{+0.07\times10^{-18}}}

\newcommand{\tidalCorrAbs}{1.9}
\newcommand{\tidalNegSeg}{12}
\newcommand{\tidalAmpNewE}{0.87}
\newcommand{\tidalSigmaNewE}{5.7}
\newcommand{\tidalAmpNewUnc}{0.15}
\newcommand{\tidalAmpNewUncE}{15}
\newcommand{\nSeg}{16}
\newcommand{\nSegAll}{17}
\newcommand{\nSamples}{986\thinspace416}

\newcommand{\nHours}{274}

\newcommand{\uSrE}{0.92}
\newcommand{\uYbE}{1.3}
\newcommand{\uLinkE}{0.1}
\newcommand{\uCombE}{0.1}
\newcommand{\dWVal}{280.042}
\newcommand{\dWUnc}{0.085}
\newcommand{\dWAlt}{280.60}
\newcommand{\dWAltUnc}{2.558}
\newcommand{\refNistV}{1.207\thinspace507\thinspace039\thinspace343\thinspace337\thinspace723\thinspace0}
\newcommand{\refNistU}{37}

\newcommand{\refEuV}{1.207\thinspace507\thinspace039\thinspace343\thinspace337\thinspace718}
\newcommand{\refEuU}{32}
\newcommand{\refEuE}{3.2}
\newcommand{\fibreKm}{1\thinspace350}

\newcommand{\siteKm}{700}
\newcommand{\levelKm}{1\thinspace113}
\newcommand{\nDays}{58}

\newcommand{\refEuBestU}{\ensuremath{7.7\times10^{-18}}}

\newcommand{\fFifteenTHz}{193.3992}
\newcommand{\deltaGV}{\ensuremath{-3.116\times10^{-15}}}
\newcommand{\tidalDiurnalPeakE}{\ensuremath{6\times10^{-18}}}
\newcommand{\tidalPpE}{1.66}
\newcommand{\uSubsystemBound}{\ensuremath{1\times10^{-19}}}
\newcommand{\levelMm}{0.255}
\newcommand{\refThreshold}{\ensuremath{5\times10^{-18}}}
\newcommand{\selfEvalBound}{\ensuremath{2\times10^{-18}}}

\newcommand{\uStaticE}{0.946}
\newcommand{\uTotalE}{2.18}
\newcommand{\uTotalShort}{2.2}
\newcommand{\RheadlineUnc}{26}
\newcommand{\dVsNistE}{\ensuremath{-5.78\times10^{-18}}}
\newcommand{\dVsEuE}{\ensuremath{-1.64\times10^{-18}}}

\newcommand{\dW}{\Delta W}

\title{Yb/Sr optical clock comparison at
\uTotalShort$\times10^{-18}$ uncertainty over a \fibreKm-km fiber link}

\title{Yb/Sr optical clock comparison at 2.2$\times10^{-18}$ uncertainty over a 1350-km fiber link}
\author{%
Lei Hou and Xiang-Pei Liu and Zhuan-Xian Xiong and Fa-Xi Chen and
Wen-Xuan Song and Jian-Yu Guan and Qiang Zhu and Yu-Chen Zhang and
Xiang Zhang and Bing Wang and De-Zhi Xiong and De-Quan Kong and
Jia-Hao Zhang and Long-Fei Liao and Shao-Mao Wang and Xing-Yang Cui and
Ping Xu and Peng Liu and Jie Li and Xiao Jiang and Ling-Xiang He and
Li-Bo Li and Qi Shen and Youchao Xie and Lei Wang and Sheng-Kai Liao and
Juan Yin and Cheng-Zhi Peng and Wen-Bin Shen and Bao-Long Lu and
Han-Ning Dai and Yu-Ao Chen and Qiang Zhang and Hai-Feng Jiang and
Jian-Wei Pan
}
\date{}

\begin{document}
\begin{center}
{\Large\bfseries Yb/Sr optical clock comparison at
2.2$\times10^{-18}$ uncertainty over a 1350-km fiber link}\par
\vspace{1.2em}
{\small Lei Hou\textsuperscript{1,2}, Xiang-Pei Liu\textsuperscript{1,3}, Zhuan-Xian Xiong\textsuperscript{4}, Fa-Xi Chen\textsuperscript{1,5}, Wen-Xuan Song\textsuperscript{1,2,3}, Jian-Yu Guan\textsuperscript{1,2}, Qiang Zhu\textsuperscript{4}, Yu-Chen Zhang\textsuperscript{1,2,3}, Xiang Zhang\textsuperscript{1,2,3}, Bing Wang\textsuperscript{4}, De-Zhi Xiong\textsuperscript{4}, De-Quan Kong\textsuperscript{1,2,3}, Jia-Hao Zhang\textsuperscript{1,2,3}, Long-Fei Liao\textsuperscript{1,2,3}, Shao-Mao Wang\textsuperscript{1,2,3}, Xing-Yang Cui\textsuperscript{1,2}, Ping Xu\textsuperscript{1,2,3}, Peng Liu\textsuperscript{1,2}, Jie Li\textsuperscript{1,2}, Xiao Jiang\textsuperscript{1,2,3}, Ling-Xiang He\textsuperscript{4}, Li-Bo Li\textsuperscript{5}, Qi Shen\textsuperscript{1,2,3}, Youchao Xie\textsuperscript{6}, Lei Wang\textsuperscript{6}, Sheng-Kai Liao\textsuperscript{1,2,3}, Juan Yin\textsuperscript{1,2,3}, Cheng-Zhi Peng\textsuperscript{1,2,3}, Wen-Bin Shen\textsuperscript{6}, Bao-Long L\"{u}\textsuperscript{4}, Han-Ning Dai\textsuperscript{1,2,3}, Yu-Ao Chen\textsuperscript{1,2,3}, Qiang Zhang\textsuperscript{1,2,3,5}, Hai-Feng Jiang\textsuperscript{1,2,3}, Jian-Wei Pan\textsuperscript{1,2,3}\par}
\vspace{0.8em}
\parbox{0.9\linewidth}{
\textsuperscript{1}Hefei National Laboratory, University of Science and Technology of China, Hefei, China\par
\textsuperscript{2}Shanghai Research Center for Quantum Sciences and CAS Center for Excellence in Quantum Information and Quantum Physics, University of Science and Technology of China, Shanghai, China\par
\textsuperscript{3}Hefei National Research Center for Physical Sciences at the Microscale and School of Physical Sciences, University of Science and Technology of China, Hefei, China\par
\textsuperscript{4}Innovation Academy for Precision Measurement Science and Technology, Chinese Academy of Sciences, Wuhan 430071, China\par
\textsuperscript{5}Jinan Institute of Quantum Technology and CAS Center for Excellence in Quantum Information and Quantum Physics, University of Science and Technology of China, Jinan 250101, China\par
\textsuperscript{6}Time and Frequency Geodesy Center (TFGC), School of Geodesy and Geomatics, Wuhan University, Wuhan 430079, China
}
\end{center}

\begin{abstract}
\sloppy
Redefining the International System of Units (SI) second requires
independent frequency-ratio measurements between clocks based on
different atomic transitions in different laboratories, with an
overall uncertainty below \refThreshold{}, to validate individual
clock evaluations through consistency checks. Reaching this
uncertainty over long distances has remained challenging. So far,
comparisons at this level have been achieved only a few times,
between co-located clocks, and their results disagree markedly.
Previous high-precision comparisons have therefore been limited to
local or kilometre-scale links. Here we compare a ytterbium optical
lattice clock in Wuhan with a strontium optical lattice clock in
Shanghai over a phase-stabilized fibre link spanning
$\fibreKm$\,km, and determine the Yb/Sr frequency ratio to be
\Rheadline(\RheadlineUnc), with a Bayesian statistical uncertainty
of \uStatShort$\times10^{-18}$ and a total uncertainty of
\uTotalE$\times10^{-18}$. To our knowledge, this is the first
intercity long distance frequency-ratio measurement whose total
uncertainty reaches the level required by the redefinition roadmap.
The measured ratio exhibits a significant correlation with Earth
tides, and its response agrees with the model within the statistical
uncertainty. At intercity scale, this comparison demonstrates the
capability of an optical-clock network for frequency comparison and
Earth's gravitational-potential sensing, a key step toward a
wide-area optical-clock network.
\end{abstract}

\section*{Introduction}
\label{sec:intro}
Optical lattice and single-ion clocks now reach fractional systematic
uncertainties at or below \selfEvalBound~\citep{mcgrew2018geodesy,
aeppli2024sr, lu2025ntsc, marshall2025alplus, jia2026sr, arnold2026lu,
zhang2026ca, zhu2026yb}, based on self-evaluations, two orders of
magnitude better than current primary frequency standards~\citep{heavner2014nistf2,
weyers2018ptbcsf}. With a new definition of the SI second on the
agenda, validating these self-assessed precision budgets is urgent.
Mandatory requirement I.2 in the roadmap for the
redefinition~\citep{dimarcq2024roadmap} calls for this task. It
requires frequency-ratio
measurements between standards of different transitions in different
laboratories to agree at an overall uncertainty
$\Delta\nu/\nu < \refThreshold$. Such agreement has yet to be
demonstrated, however. Resolution~4 of the 28th General Conference
on Weights and Measures (CGPM)
(2026)~\citep{cgpm2026resolution4},
while maintaining the plan to present a proposal for the new
definition of the second at the 29th CGPM (2030), notes
that ``there are noticeable inconsistencies in some
comparisons of optical frequency standards and in the global set of
frequency ratio measurements involving OFS''.

Measurements at this uncertainty level have so far been realized only
locally, a few times~\citep{beloy2021bacon, takamoto2020redshift,
aeppli2026nist}. The highest-performance
clocks are complex, bulky systems that are difficult to transport, so
linking clocks of different types and locations into a
collaborative network is a key pathway toward establishing
optical clocks as a verifiable metrological
infrastructure~\citep{komar2014quantumnetwork, riehle2017networks}. The
idea has moved from proposal to practice. Fiber links with noise levels
below those of the clocks themselves now span thousands of
kilometers~\citep{schioppo2022cavities, chen2026dissemination}, and
multi-clock campaigns have measured tens of ratios among clocks in six
countries~\citep{lindvall2025coordinated} and across a continental
network~\citep{pizzocaro2026european}. Satellite-based advanced links
with comparable performance~\citep{shen2021satellite, shen2022freespace,
caldwell2023geo, derevianko2022spaceclock} are also being developed and
will interconnect with fiber networks to form a future global optical
clock network. However, comparisons across independent institutions
and different locations have reached at best a precision of
\refEuBestU~\citep{pizzocaro2026european}; so far, no frequency-ratio
measurement between clocks in different cities has satisfied this
uncertainty requirement~\citep{fortier2026review}.

Closing this comparison gap matters beyond the definition itself.
Chronometrically
leveled clocks map the gravitational potential of the Earth at
centimeter resolution~\citep{grotti2018geodesy, mcgrew2018geodesy,
takamoto2020redshift, bothwell2022millimetre}, and clock networks
have been proposed and used as
sensors for dark matter~\citep{derevianko2014topological,
wcislo2018global, beloy2021bacon}.

Here we report an intercity frequency-ratio measurement
between a ytterbium lattice clock at the Innovation Academy for
Precision Measurement Science and Technology (APM) in Wuhan and a
strontium lattice clock at the University of Science and Technology
of China (USTC) in Shanghai, about $\siteKm$\,km apart. We determine
the Yb/Sr frequency ratio to
be \Rheadline(\RheadlineUnc), with the lowest total uncertainty
reported to date for an independent-clock comparison,
\uTotalE$\times10^{-18}$.

\section*{Results}
\label{sec:results}
\subsection*{A two-node optical-clock network}
\label{sec:results-network}

\begin{figure}[t]
  \centering
  \includegraphics[width=\linewidth]{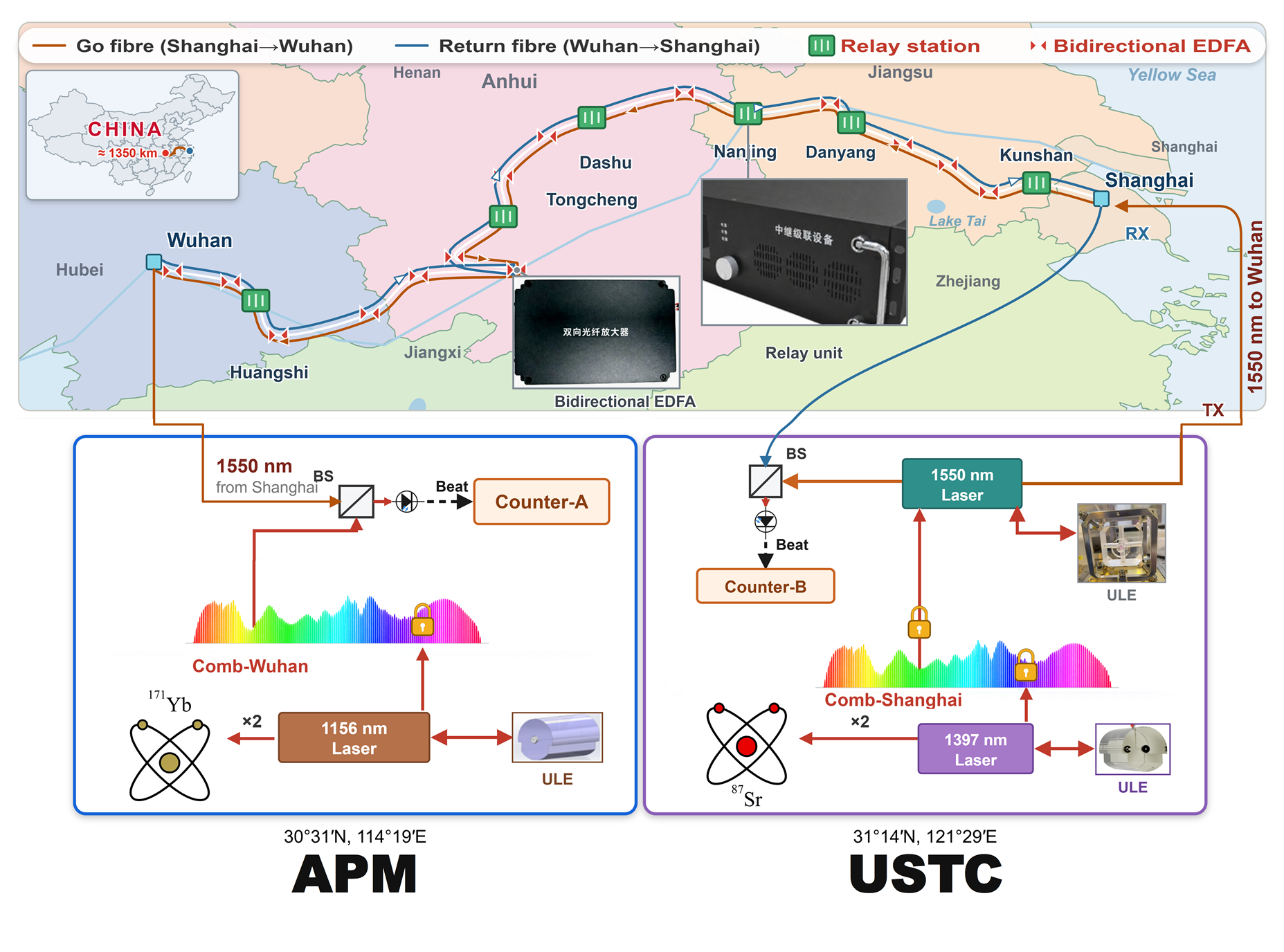}
  \caption{The two-clock network. A ytterbium clock at APM in Wuhan and
  a strontium clock at USTC in Shanghai are connected by a
  phase-stabilized
  \qty{1550}{nm} fiber link. Relay stations along the route carry
  erbium-doped fiber amplifiers and fiber-noise-cancellation loops.
  ULE, ultra-low-expansion glass optical cavity; BS, beam
  splitter.}
  \label{fig:network}
\end{figure}

Figure~\ref{fig:network} shows the two-clock network. A ytterbium (Yb)
lattice clock at APM (the Innovation Academy for Precision Measurement
Science and Technology)~\citep{zhu2026yb} in Wuhan
(\ang{30;31;} N, \ang{114;19;} E) and a strontium (Sr)
lattice clock at USTC (the University of Science and Technology of
China)~\citep{jia2026sr} in Shanghai
(\ang{31;14;} N, \ang{121;29;} E) are
compared through a $\fibreKm$\,km
phase-stabilized fiber link on a commercial telecom network used
non-exclusively, with the carrier transmitted from Shanghai to Wuhan;
the route is carried by 21 relay stations with 22 erbium-doped fiber
amplifiers in total. The frequency-ratio measurement is
obtained from the beat between the received carrier $f_{1550}$ and
the local Yb-referenced comb, recorded once per second by frequency
counter~A, while counter~B continuously monitors the out-of-loop
noise of the round-trip fiber link in real time; all optical
phase-locked loops in the chain are likewise
monitored to evaluate the status of the experiment. The geopotential
difference between the two clocks was
measured by leveling to be
$\dW = \dWVal \pm \dWUnc~\mathrm{m^{2}\,s^{-2}}$ (Methods).

\subsection*{Clock-ratio determination}
\label{sec:results-ratio}

\begin{figure}[t]
  \centering
  \includegraphics[width=\linewidth]{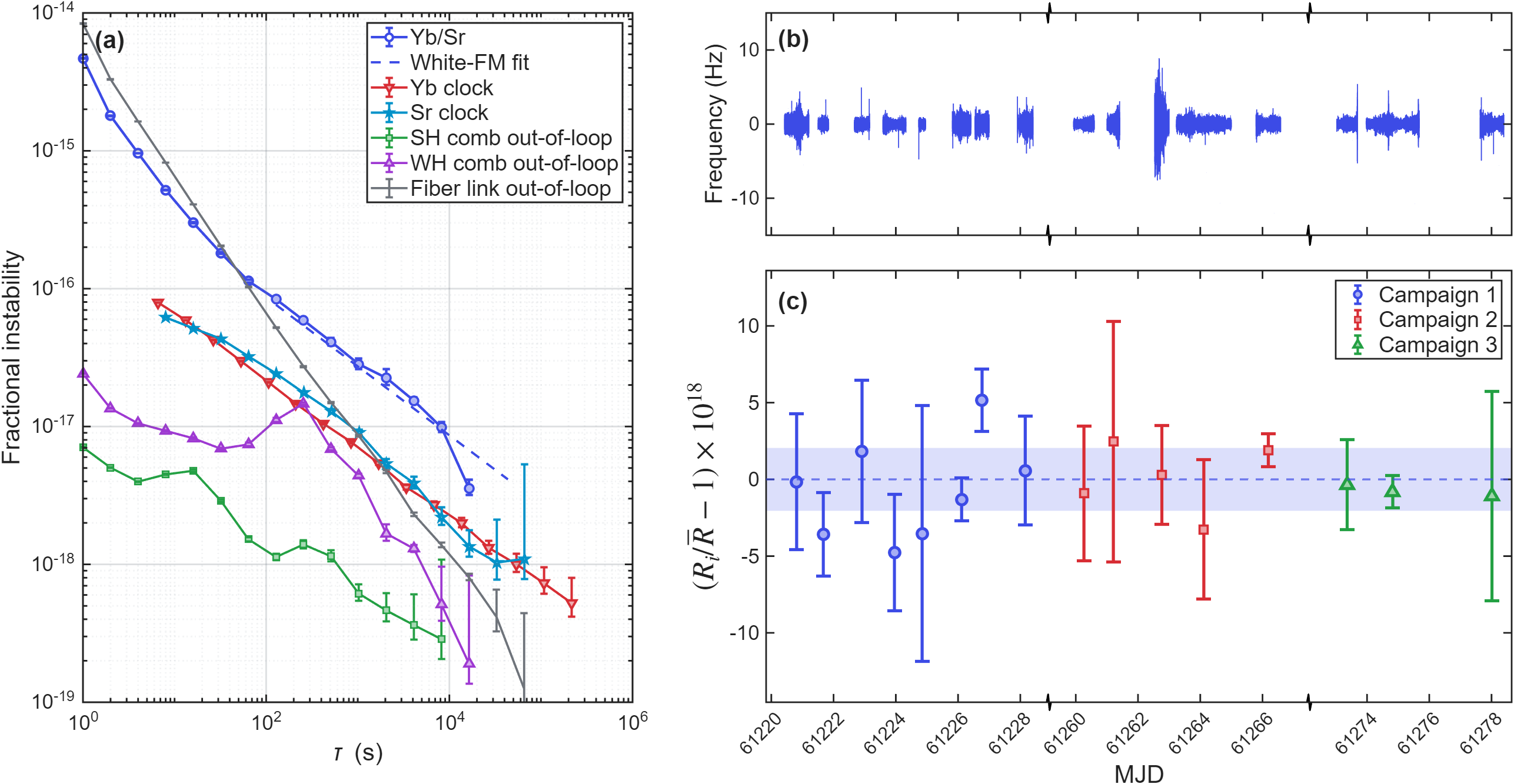}
  \caption{Frequency-ratio measurement results. (a) Fractional
  instability of a representative segment (Yb/Sr), together with the
  instabilities of the Yb and Sr clocks and the
  two frequency combs (SH, WH), all measured beforehand,
  and the real-time out-of-loop
  monitor of the fiber link, and the white-frequency-noise fit.
  (b)
  The raw beat record from frequency counter A, showing measurement
  times and beat-frequency fluctuations around its mean offset.
  (c) Per-segment Yb/Sr ratio deviations from the final reported
  value with their statistical uncertainties, colored by measurement
  campaign.}
  \label{fig:ratio}
\end{figure}

Over two months (\nDays~days) the two clocks and their transfer chains
ran synchronously for a total of \nHours~effective hours, recorded as
\nSeg~jump-free segments in three measurement stages defined by the
experiment team (late June to early July, early August, and late
August); several of the segment boundaries
correspond to deliberate clock shutdowns and restarts, so the
reproducibility of the ratio rests on tested restart behavior rather
than on continuous operation. One further data-taking period, run
while the experimental parameters were still being tuned, is treated
as preparation and testing and is not included in the combination;
its segment shows anomalous values of several operating parameters
and of the resulting ratio, as does the \nSegAll-segment combination
that includes it (Supplementary Section~1). Figure~\ref{fig:ratio}a shows
the instabilities of the Yb/Sr ratio and of the round-trip fiber
link within a representative segment, together with the clock and
comb instabilities measured beforehand. At short
averaging times the fiber link dominates, because its length limits
the feedback bandwidth and thus its short-term stability; at long
averaging times the combined Yb/Sr instability lies above the
combined instability of the two clocks, revealing other noise
contributions, including the time-varying gravitational redshift
discussed in the Discussion. Figure~\ref{fig:ratio}b displays the
actual measurement times and the fluctuations of the beat record
around its large mean offset, which stay within $\pm 10$\,Hz.

We determine the Yb/Sr frequency ratio from the \nSeg~segments. Each
segment yields a ratio estimate with its own statistical uncertainty;
Figure~\ref{fig:ratio}c shows each segment's ratio deviation with its
uncertainty, colored by measurement campaign, and a
straight-line fit to the segment values against time gives a drift
rate of $\dDriftPerDay$ per day, $0.4\sigma$ away from zero.

For each segment, the statistical uncertainty is
taken as the last computable point of the overlapping Allan deviation
rather than an extrapolation of the white-frequency slope. In a
remote comparison the extra noise of a ratio measurement between
clocks at two separated sites cannot simply be expected to follow a
white-frequency-noise pattern, because other effects beyond the clock
contributions enter the long-term record, as already visible in
Figure~\ref{fig:ratio}a. A direct comparison of the two choices shows
that the extrapolation underestimates the last-point stability by
30--40\% on average.

Following recent ratio
comparisons~\citep{pizzocaro2026european, aeppli2026nist}, we combine
the segments with a random-effect model that admits a common excess
scatter~\citep{birge1932errors, mandel1970interlaboratory}, and
evaluate three estimators (Methods); the three
center values agree to within $1\times10^{-20}$, giving \RrawHead. Accounting for
the time-varying gravitational redshift as discussed in the
Discussion, the Yb/Sr ratio is determined as
\Rheadline.

\subsection*{Uncertainty budget}
\label{sec:results-budget}

Table~\ref{tab:budget} is the uncertainty budget of the Yb/Sr
frequency-ratio measurement.

The statistical uncertainty of the ratio combination is
$\uStatE\times10^{-18}$, taken from the Bayesian random-effect
estimator, the one with the largest statistical uncertainty of the
three. The Sr clock contributes $\uSrE\times10^{-18}$ from its
published systematic evaluation~\citep{jia2026sr}, and the Yb clock
$\uYbE\times10^{-18}$~\citep{zhu2026yb}. The static
gravitational-potential term is the leveling uncertainty of $\dW$,
$\dWUnc~\mathrm{m^2\,s^{-2}}$, converted to fractional frequency
by division by $c^2$ with $c$ the speed of light, giving
$\uStaticE\times10^{-18}$. The fiber link and
the frequency comb each contribute below $\uSubsystemBound$, from the
digitization error of the direct-digital-synthesis
frequency-reference settings, and are carried as \qty{0.1e-18} upper
bounds in the quadrature.

Combining these components in quadrature gives the total standard
uncertainty $\uTotalE\times10^{-18}$; the reported ratio is therefore
\begin{equation}
  \mathrm{Yb/Sr} = \Rheadline(\RheadlineUnc),
\end{equation}
where the parenthesized digits are the total standard uncertainty in
the last two decimal places of the value.

\begin{table}[t]
  \centering
  \caption{Uncertainty budget of the Yb/Sr frequency-ratio
  measurement. Components are treated as independent; the fiber-link
  and frequency-comb entries are upper bounds (each below
  $\uSubsystemBound$, from the digitization error of the
  direct-digital-synthesis frequency-reference settings), carried at
  \qty{0.1e-18} in the quadrature.}
  \label{tab:budget}
  \begin{tabular}{lS[table-format=<1.3]}
    \toprule
    Component & {Contribution ($10^{-18}$)} \\
    \midrule
    Statistical (Bayesian random effect) & \uStatE \\
    Sr systematics                       & \uSrE \\
    Yb systematics                       & \uYbE \\
    Static gravitational potential       & \uStaticE \\
    Fiber link                           & <\uLinkE \\
    Frequency comb                       & <\uCombE \\
    \midrule
    Total (quadrature)                   & \uTotalE \\
    \bottomrule
  \end{tabular}
\end{table}

\section*{Discussion and outlook}
\label{sec:discussion}
Unlike a co-located comparison, a remote one is affected not only by
the static gravitational-potential difference, corrected once from
the leveled geopotential (Methods), but also by the time-varying
gravitational redshift, whose diurnal component reaches up to
\tidalDiurnalPeakE over the
campaign, against a total peak-to-peak variation of
$\tidalPpE\times10^{-17}$ (Supplementary Section~4). The beat
record carries a signal at the tidal period at the $10^{-18}$ level.
No single segment resolves it on its own, but the \nSeg~segments
combined leave no doubt (Methods, Supplementary Figure~2). The
within-segment correlation against a
template built from the DE421
ephemerides~\citep{folkner2009de421} and the FES2022 ocean-tide
model~\citep{lyard2021fes2014} is positive in \tidalNegSeg~out of
\nSeg~segments, and the precision-weighted response amplitude is
$\tidalAmpNewE\,(\tidalAmpNewUncE)$ ($1\sigma$), which differs from
zero at $\tidalSigmaNewE\sigma$ and agrees with the expected
response of $+1$ within the statistical uncertainty. The fitted
response thus validates the template: since the measured beat
reproduces the predicted waveform to within the statistical
uncertainty, the model is adopted as the best available description
of the time-varying term, and its theoretical amplitude is applied
as a bias correction to the central value (Methods). The correction
is dominated by the campaign mean of the template and shifts the
reported ratio by $\tidalCorrAbs\times10^{-19}$, a small quantity
compared with the total uncertainty.

\begin{figure}[t]
  \centering
  \includegraphics[width=\linewidth]{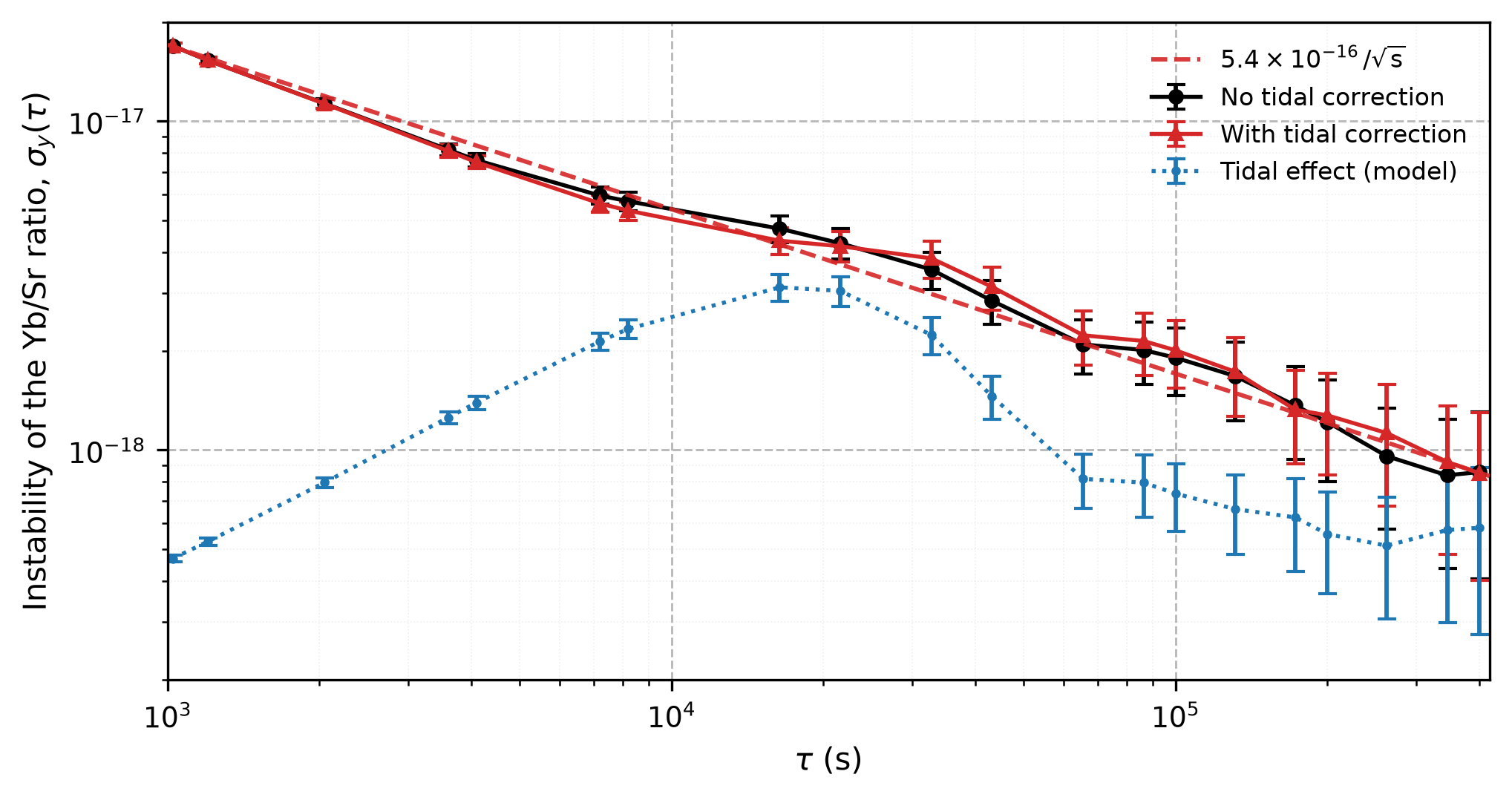}
  \caption{Spliced concatenated instability of the \nSeg~segments
  (gaps removed, $\nSamples$ samples). One overlapping Allan
  deviation per series, $\tau$ from \qty{1000}{s} to \qty{4.6}{d}.
  The three curves show the uncorrected record, the DE421+FES2022
  full-tide correction ($A=+1$) and the removed tidal correction
  term (dotted). The red dashed line marks
  $5.4\times10^{-16}\,/\sqrt{\mathrm{s}}$.}
  \label{fig:spliced}
\end{figure}

Figure~\ref{fig:spliced} shows the spliced instability of the
corrected beat data. At averaging times of a few hours, where the
tidal effect is largest, the correction
improves the record, confirming that the adopted model captures the
dominant part of the time-varying redshift. In the diurnal and
semi-diurnal
ocean-loading band it degrades it, indicating that the ocean-loading
model is the weaker part. At longer averaging times the periodic tide
averages out, leaving only the campaign mean, which in our case is
more than an order of magnitude smaller than the dynamic range, so
the correction has little effect there. At
the longest averaging times the record reaches
$8.5\times10^{-19}$ at 4.6\,d, and the overall
trend follows the white-frequency-noise slope,
$\sigma_y(\tau) = 5.4\times10^{-16}/\sqrt{\tau/\mathrm{s}}$, in
line with the statistical combination, whose Bayesian uncertainty
of $\uStatE\times10^{-18}$ (Table~1) follows from the conservative
per-segment last-point Allan deviations.

The tidal effect grows with the scale of the comparison. On
continental- and intercontinental-scale baselines the time-varying
redshift becomes one of the dominant systematic
terms~\citep{denker2018geodetic}. That same sensitivity turns the
network into an instrument for the Earth. Clock comparisons over
large baselines directly sense the time-variable gravitational
potential, providing an independent means of testing and calibrating
tide and loading models at the $10^{-18}$ level, where the limits of
the current model seen here can be refined, or removed, by more
comparison data from an expanding optical-clock network. Conversely,
tidal knowledge gained in this way feeds back into the comparisons
themselves. With data-consistent models, the time-varying term can
be corrected at the $10^{-19}$ level, extending the accuracy reach
of large-scale clock comparisons.

\begin{figure}[t]
  \centering
  \includegraphics[width=\linewidth]{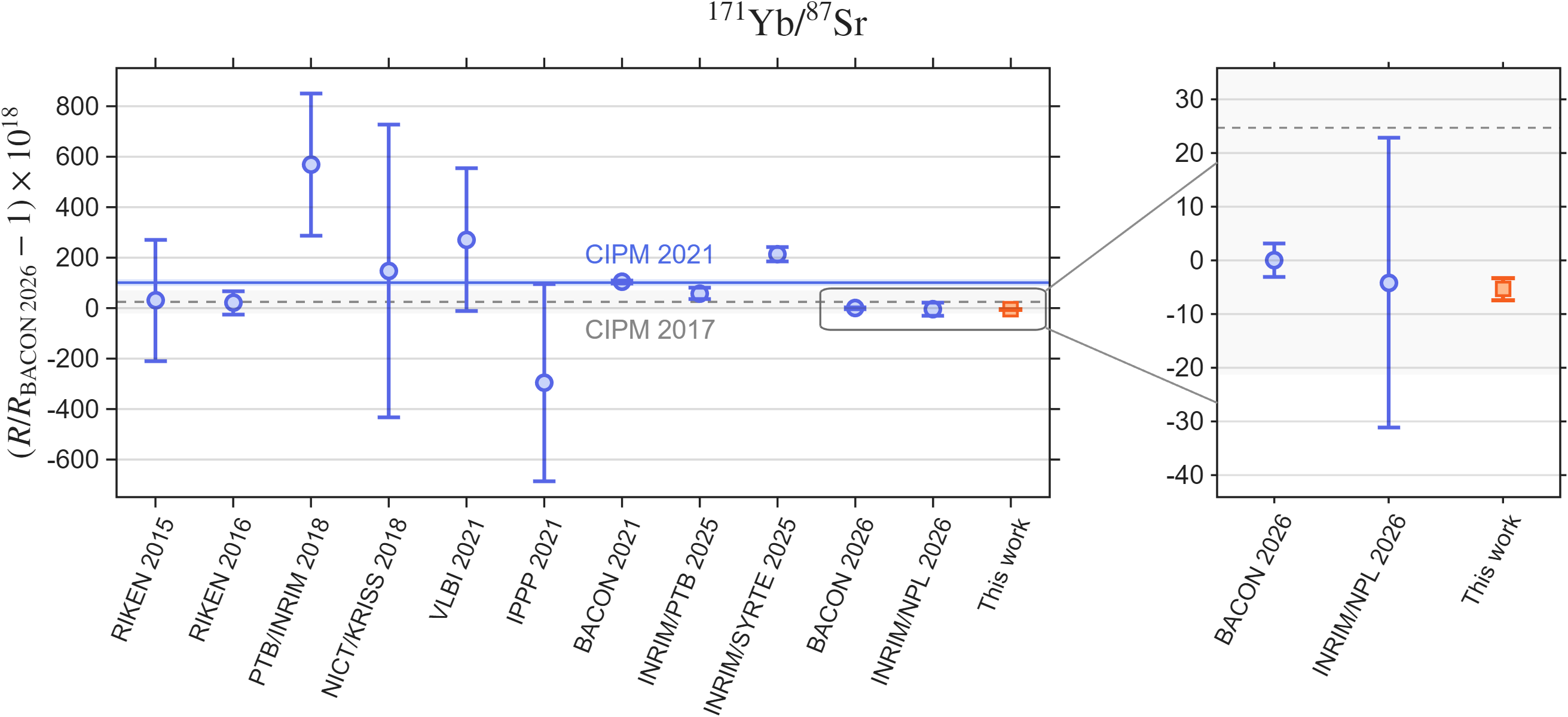}
  \caption{Evolution of the Yb/Sr frequency-ratio measurements.
  Deviations from the final value are shown against the measurement
  year for the historical determinations and for this work.}
  \label{fig:history}
\end{figure}

The evolution of the Yb/Sr ratio shows how comparison measurements
drive the field forward. Since
2016~\citep{nemitz2016ybsr}, independent ratios were measured with
increasing precision, yet the determinations disagreed well beyond
their quoted uncertainties, an early sign of common-mode errors in
the absolute frequency chain on which the earlier values
relied~\citep{beloy2021bacon}. The successive comparisons exposed
those errors, guided their elimination, and fed the corrected
evaluations back into the participating laboratories, so that the
determinations have converged steadily
(Figure~\ref{fig:history}).

The value reported here differs from the NIST--JILA measurement by
$\dVsNistE$ relative to the NIST--JILA value
\refNistV(\refNistU), beyond the combined
standard uncertainty of the two measurements; the gap, however, is
now far smaller than the few $\times10^{-16}$ disagreements seen
only a few years ago~\citep{aeppli2026nist}, and
the recent trend is one of gradual convergence. The
European-network value
\refEuV(\refEuU)~\citep{pizzocaro2026european}, which differs from
ours by $\dVsEuE$ at a larger uncertainty of
\refEuE$\times10^{-17}$, agrees with ours within its uncertainty and
lies between the two.

The established frequency-comb technology and a fiber link with
out-of-loop monitoring are each insufficient to produce a ratio bias
at such a level. The most plausible remaining origin is the
systematic evaluation of the clocks, which will be
investigated in dedicated follow-up measurements. To date, no
frequency ratio has yet been confirmed by two independent
measurements at the level required for the redefinition. Closing the
remaining gap will require more high-performance clocks to join the
ratio-measurement effort, and the network to be established
promptly, so that the second redefinition targeted for 2030/2034 can
be realized.

Looking ahead, the result is a milestone on the necessary path
toward a wide-area optical-clock network. The building blocks are now
in place. Terrestrial fiber links of the
kind used here~\citep{schioppo2022cavities, chen2026dissemination}
carry comparisons over thousands of kilometers within continents,
while free-space and satellite-based links are closing the remaining
gaps~\citep{pizzocaro2021vlbi, shen2021satellite,
shen2022freespace, caldwell2023geo}. The natural next step is a
hybrid architecture in which optical-clock nodes on continental fiber
backbones are interconnected across oceans by satellite or free-space
optical links, so that any pair of clocks on Earth can be compared at
the level demonstrated here, a network that would also multiply the
diagnostic opportunities for resolving discrepancies such as the one
reported above. Such a network and the geodetic
infrastructure will develop together. Precise geopotential
referencing supports the comparisons, while the network's
time-variable gravity observations in turn refine the geodetic
models~\citep{grotti2018geodesy, mcgrew2018geodesy}.

\section*{Conclusion}
\label{sec:conclusion}
We have compared an ytterbium lattice clock in Wuhan and a strontium
lattice clock in Shanghai over a \fibreKm\,km fiber link and
determined the Yb/Sr frequency ratio with a total standard
uncertainty of \uTotalE$\times10^{-18}$, the lowest reported to date
for an independent-clock comparison. The beat record responds to the
tides and agrees with the model response within the statistical
uncertainty, so the comparison directly senses the time-varying
gravitational potential across the \siteKm\,km baseline. The
remaining disagreement with the NIST--JILA determination keeps the
accuracy budgets of the participating clocks open questions, which
the redefinition of the second requires to be closed. The two
sites now form a working node pair of a distributed
optical-clock network. With
more optical-clock nodes joined through terrestrial fiber backbones
and interconnected across continents by satellite and free-space
links, the comparison demonstrated here extends naturally into the
global network on which the redefinition, chronometric leveling of
the gravitational potential, and distributed searches for new physics
will depend, and on which disagreements between independent clocks
can be identified and resolved.



\section*{Methods}
\subsection*{Leveling determination of the geopotential difference}

The static gravitational-redshift correction is fixed by the
geopotential difference between the atomic-transition positions of the
two clocks, determined by a first-order leveling campaign
carried out specifically for this comparison: a
$\levelKm$\,km line between the Wuhan benchmark WS-03 and the
Shanghai benchmark WS-01 through Ezhou, Huangshi, Huanggang, Lu'an,
Hefei, Wuhu, Guangde and Suzhou, observed in both directions with
digital levels and bar-code invar staffs, with sight lengths not
exceeding \qty{30}{m}, the line checked in the field in
\qtyrange{6}{8}{km} sections, and gravity observed along the line
with two relative gravimeters. The geopotential difference is
integrated section by section,
$\dW_{AB} = -\sum_i g_i\,\Delta H_i$, and combined with the local
benchmark-to-transition offsets, giving
$\dW = \dWVal \pm \dWUnc~\mathrm{m^2\,s^{-2}}$, equivalent to an
optical fractional shift of $\deltaGV$. The per-kilometer accidental
error $M_\Delta = \levelMm$\,mm\,km$^{-1}$ satisfies the most
stringent first-order requirements of the Chinese (GB/T 12897--2006,
$M_\Delta \le 0.45$\,mm\,km$^{-1}$), US (FGCC 1984,
$0.5$\,mm\,km$^{-1/2}$), German (AdV FA-PN 2022, $0.4$\,mm) and
Czech (\v{C}\'UZK 2015, $m_0 = 0.40$\,mm\,km$^{-1}$) national
standards~\citep{fgcc1984,gb12897,cuzk2015,advfapn2022}. An
independent GNSS-plus-geoid check used precise point positioning for
the ellipsoidal heights of the permanent marks, local leveling for
the vertical offsets to the clock reference surfaces, and the EGM2008
geoid model for the conversion to geopotential, giving
$\dWAlt \pm \dWAltUnc~\mathrm{m^2\,s^{-2}}$, consistent with the
leveled value but about 30 times less precise, so it serves as a
consistency check rather than a calibration at the reported
accuracy.

\subsection*{Frequency transfer}

The two clocks are linked by a phase-stabilized fiber network. At
each end an optical
frequency comb, self-referenced and phase-locked to the local clock,
bridges the clock wavelength to the \qty{1550}{nm} telecom carrier;
all frequencies in the chain thus derive from the two clock
transitions through fixed factor-of-two relations, so the two clocks
are the only independent frequency sources in the system
(Supplementary Figure~1).

The frequency relations at the Shanghai end are
\begin{align}
  f_{1397} &= \tfrac{1}{2}(1+a)\,f_{\rm Sr}
           = N_{\rm Sr}\,f_{\rm rep1} + f_{\rm ceo1} - f_{\rm beat1}, \\
  f_{1550} &= N_{1550}^{\rm SH}\,f_{\rm rep1} + f_{\rm ceo1} + f_{\rm beat2},
\end{align}
where $f_{1397}$ is the frequency of the \qty{1397}{nm} light
obtained by doubling the \qty{698}{nm} Sr clock light,
$f_{\rm Sr}$ the Sr clock
transition frequency, $N_{\rm Sr}$ the tooth count at which the
light is locked to the Shanghai comb, and $f_{\rm rep1} =
\qty{250}{MHz}$, $f_{\rm ceo1}$ and $f_{\rm beat1}$ the repetition
rate, carrier-envelope offset frequency and lock beat of that comb;
$f_{1550}$ is the \qty{1550}{nm} transfer carrier, $N_{1550}^{\rm
SH}$ its tooth count and $f_{\rm beat2}$ its beat against the comb,
and $a$ collects the Sr corrections. The carrier at
$f_{1550} + f_{\rm rep1}/25$ is transmitted to Wuhan. The
corresponding relations at the Wuhan end are
\begin{align}
  f_{1156} &= \tfrac{1}{2}(1+b)\,f_{\rm Yb}
           = N_{\rm Yb}\,f_{\rm rep2} + f_{\rm ceo2} - f_{\rm beat3}, \\
  f_{1550} + f_{\rm rep1}/25 &= N_{1550}^{\rm WH}\,f_{\rm rep2} + f_{\rm ceo2} + f_{\rm beat4},
\end{align}
where $f_{1156}$ is the frequency of the \qty{1156}{nm} light
obtained by doubling the \qty{578}{nm} Yb clock light,
$f_{\rm Yb}$ the Yb clock
transition frequency, $N_{\rm Yb}$ the tooth count at which the
light is locked to the Wuhan comb, and $f_{\rm rep2} =
\qty{200}{MHz}$, $f_{\rm ceo2}$ and $f_{\rm beat3}$ the repetition
rate, carrier-envelope offset frequency and lock beat of that comb;
$N_{1550}^{\rm WH}$ is the tooth count of the received carrier,
$f_{\rm beat4}$ its beat against the Wuhan comb, and $b$ collects
the Yb corrections. Eliminating the intermediate frequencies yields
the ratio of the two clock transitions:
\begin{equation}
  \frac{f_{\rm Yb}}{f_{\rm Sr}} =
  \frac{\dfrac{1+b}{N_{\rm Yb}}
    \left(N_{1550}^{\rm WH} + \dfrac{f_{\rm ceo2}}{f_{\rm rep2}}
    + \dfrac{f_{\rm beat4}}{f_{\rm rep2}}\right)}
  {\dfrac{1+a}{N_{\rm Sr}}
    \left(N_{1550}^{\rm SH} + \dfrac{f_{\rm ceo1} + f_{\rm beat2}}
    {f_{\rm rep1}} + \dfrac{1}{25}\right)} .
\end{equation}
The $N$ counts and the composition of $a$ and $b$ are listed in
Supplementary Section~2.

\subsection*{Data processing}

The observable is the beat between the received carrier and the
local Yb-referenced comb, sampled every second in phase-averaging
mode; the beat is recorded by frequency counter A, while counter B
continuously monitors the out-of-loop noise of the round-trip fiber
link. The beat is converted to fractional units with the transfer
frequency $f_{1550} = \fFifteenTHz$\,THz. The static
gravitational-redshift correction enters the ratio inversion as a
multiplicative factor,
$\left(\mathrm{Yb/Sr}\right)_{\rm corr} =
\left(\mathrm{Yb/Sr}\right)_{\rm raw} \times (1+\delta_g)$ with
$\delta_g = \deltaGV$. Each accepted segment contributes its longest
jump-free record; the per-segment retained counts, ratio deviations
and statistical uncertainties are listed in Supplementary Table~1.
For each segment the statistical uncertainty
$u_i = R_i\,\sigma_y(\tau_{\max})$ adopts the last computable point
of the overlapping Allan deviation $\sigma_y(\tau)$, the two-sample
deviation of contiguous $\tau$-averages of the record evaluated
with full overlap, on a tau grid of powers
of two plus 600, 1200, 3600 and 7200\,s and limited to a quarter of
the longest gap-free run; extrapolating the white-frequency slope
instead would systematically underestimate $u_i$
(Supplementary Section~3). Five segments whose stability slopes fall
outside the white-frequency-noise band
are flagged there as model-limited; they lie at the steep end of a
continuous slope distribution with no common campaign, weather or
laboratory condition that distinguishes them from the remaining
segments.

\subsection*{Statistical combination}

The segment ratios are combined under a random-effect model with
common excess scatter, $R_i \sim \mathcal{N}(\mu, u_i^2 + \xi^2)$.
Three estimators are evaluated as cross-checks. The weighted least
squares uses inverse-variance weights $w_i = 1/u_i^2$; the
Mandel--Paule estimator chooses $\xi$ so that
$\chi^2_{\rm red}(\xi) = 1$; the Bayesian random effect places flat
priors on $\mu$ and $\log\xi$ and marginalizes the joint posterior
numerically over a 40-decade $\log\xi$ grid (401 points, trapezoidal
quadrature), expanding the span until the posterior mass at the
edges falls below 1\%. The Bayesian posterior mean is adopted for the
reported value; the estimator formulas and the $\log\xi$ grid
details are in Supplementary Section~3.

\begin{table}[t]
  \centering
  \caption{Statistical combination of the \nSeg~segment ratios.
  WLS denotes weighted least squares with inverse-variance weights
  $w_i = 1/u_i^2$; $\xi$ is the inferred common excess scatter.
  $\chi^2_{\rm red}$ is the reduced chi-square of the WLS
  combination ($\dofV$ degrees of freedom, $p = \pChi$, Birge ratio
  $\birgeRatio$).}
  \label{tab:estimators}
  \begin{tabular}{lS[table-format=1.21]S[table-format=1.3]S[table-format=1.2]S[table-format=1.3]}
    \toprule
    Estimator & {Yb/Sr} & {$u$ ($10^{-18}$)} & {$\xi$ ($10^{-18}$)} & {$\chi^2_{\rm red}$} \\
    \midrule
    WLS (inverse variance)   & \RwlsRawT & \uWlsE & {---} & \chiRed \\
    Mandel--Paule            & \RmpRawT & \uMpE & \xiMpE & 1 \\
    Bayesian random effect   & \RbayesRawT & \uStatE & \xiBayesE & {---} \\
    \bottomrule
  \end{tabular}
\end{table}

\subsection*{Tidal correlation analysis}

The tidal signal in the comparison is quantified by a
within-segment response analysis of the beat
waveform against the tide template; the spliced
instability of the concatenated record before and after the
correction is shown in the Discussion (Figure~\ref{fig:spliced}).
The per-segment numbers are listed in Supplementary
Table~5.

\emph{Tidal template.} The template construction is described in
full in Supplementary Section~4. In brief, the solid Earth tide is
computed from the DE421 ephemerides with degree-2 Love numbers, and
the ocean-tide loading correction is taken from the FES2022
loading-tide atlas; the beat-frequency template is
\[
h(t) = f_{1550}\,\Delta W(t)/c^2,
\]
where $\Delta W(t)$ is the time-varying part of the potential
difference, distinct from the static $\dW$ determined by leveling
above. The same $f_{1550}$ normalization as the beat record is used,
with the direction convention $\Delta W = W(\mathrm{Wuhan}) -
W(\mathrm{Shanghai})$.

\begin{figure}[t]
  \centering
  \includegraphics[width=\linewidth]{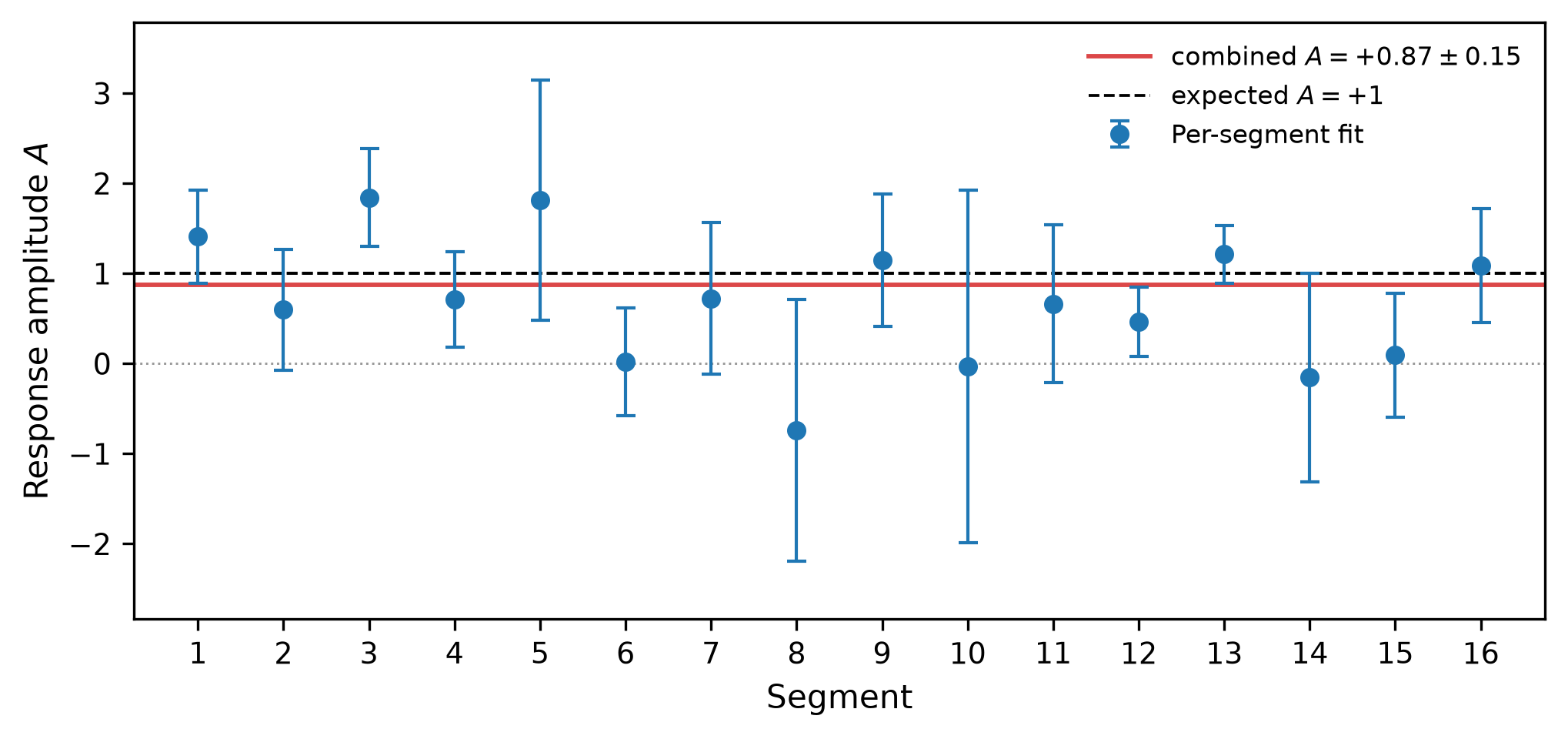}
  \caption{Per-segment tidal response amplitudes $A$ and their
  uncertainties (DE421+FES2022), 1200-s windowed, sign convention
  with the expected response at $A=+1$. The red line marks the
  precision-weighted combination $\tidalAmpNewE\,(\tidalAmpNewUncE)$,
  significant at $\tidalSigmaNewE\sigma$; no single segment is
  individually significant.}
  \label{fig:forest}
\end{figure}

\emph{Within-segment response.} The retained beat record $b(t)$ and
the 1-s template $h(t)$ of each segment are both window-averaged with
a 1200-s triangular window at 600-s stride, demeaned, and fitted
through the origin,
\[
b = A\,h + \varepsilon,
\]
with the template sign corrected for the beat polarity so that a
positive amplitude corresponds to the expected response $A = +1$.
Because the 600-s-stride windows overlap, inference uses only
non-overlapping 1200-s windows: the amplitude, the Pearson
correlation $r$ and its $p$ value are estimated there, and the
per-segment amplitude uncertainty $u_A$ is the scatter of the
non-overlapping residuals. The per-segment amplitudes are
combined with precision weights $1/u_A^2$ into the combined response,
whose significance relative to zero is $A/u_A$; the per-segment signs
are tested against the symmetric binomial null.
Figure~\ref{fig:forest} shows the per-segment fits and the combined
amplitude. As a cross-check, the same fit is
repeated with the solid-Earth tide alone, which locates the dominant
coupling in the ocean-tide loading term
(Supplementary Section~5).

\emph{Correction.} The tide series is applied as a bias correction,
$b_{\rm corr}(t) = b(t) - A\,h(t)$, at the theoretical amplitude
$A = +1$; its campaign mean is applied to the central value, and
the resulting combination statistics are listed in Supplementary
Table~7. The reported central value carries the bias correction.

\section*{Data availability}
The frozen, machine-readable result layer underlying this manuscript
(the manifest with SHA-256
\texttt{ba3c4343269182890f2f3b4b2248c651cf8cd8e1c7f646bdb4f36e5ffe2fa4bf},
the derived ledgers and statistics, and the generated audit report) is
available from the corresponding author upon reasonable request, as are
the raw beat files and the leveled geopotential inputs.

\section*{Code availability}
The analysis and manuscript pipelines are in the same repository under the
Apache License 2.0. The manuscript numbers are generated by
\texttt{paper/rev2/tools/build\_numbers.py}, which reads only the frozen
manifest and the curated constants listed in
\texttt{generated/numbers\_manifest.json}.

\clearpage
{\Large\bfseries Supplementary Information: Yb/Sr optical clock comparison
at 2.2$\times10^{-18}$ uncertainty over a 1350-km fiber link}\par
\vspace{1em}
\begin{center}
{\small Lei Hou\textsuperscript{1,2}, Xiang-Pei Liu\textsuperscript{1,3}, Zhuan-Xian Xiong\textsuperscript{4}, Fa-Xi Chen\textsuperscript{1,5}, Wen-Xuan Song\textsuperscript{1,2,3}, Jian-Yu Guan\textsuperscript{1,2}, Qiang Zhu\textsuperscript{4}, Yu-Chen Zhang\textsuperscript{1,2,3}, Xiang Zhang\textsuperscript{1,2,3}, Bing Wang\textsuperscript{4}, De-Zhi Xiong\textsuperscript{4}, De-Quan Kong\textsuperscript{1,2,3}, Jia-Hao Zhang\textsuperscript{1,2,3}, Long-Fei Liao\textsuperscript{1,2,3}, Shao-Mao Wang\textsuperscript{1,2,3}, Xing-Yang Cui\textsuperscript{1,2}, Ping Xu\textsuperscript{1,2,3}, Peng Liu\textsuperscript{1,2}, Jie Li\textsuperscript{1,2}, Xiao Jiang\textsuperscript{1,2,3}, Ling-Xiang He\textsuperscript{4}, Li-Bo Li\textsuperscript{5}, Qi Shen\textsuperscript{1,2,3}, Youchao Xie\textsuperscript{6}, Lei Wang\textsuperscript{6}, Sheng-Kai Liao\textsuperscript{1,2,3}, Juan Yin\textsuperscript{1,2,3}, Cheng-Zhi Peng\textsuperscript{1,2,3}, Wen-Bin Shen\textsuperscript{6}, Bao-Long L\"{u}\textsuperscript{4}, Han-Ning Dai\textsuperscript{1,2,3}, Yu-Ao Chen\textsuperscript{1,2,3}, Qiang Zhang\textsuperscript{1,2,3,5}, Hai-Feng Jiang\textsuperscript{1,2,3}, Jian-Wei Pan\textsuperscript{1,2,3}\par}
\vspace{0.8em}
\parbox{0.9\linewidth}{
\textsuperscript{1}Hefei National Laboratory, University of Science and Technology of China, Hefei, China\par
\textsuperscript{2}Shanghai Research Center for Quantum Sciences and CAS Center for Excellence in Quantum Information and Quantum Physics, University of Science and Technology of China, Shanghai, China\par
\textsuperscript{3}Hefei National Research Center for Physical Sciences at the Microscale and School of Physical Sciences, University of Science and Technology of China, Hefei, China\par
\textsuperscript{4}Innovation Academy for Precision Measurement Science and Technology, Chinese Academy of Sciences, Wuhan 430071, China\par
\textsuperscript{5}Jinan Institute of Quantum Technology and CAS Center for Excellence in Quantum Information and Quantum Physics, University of Science and Technology of China, Jinan 250101, China\par
\textsuperscript{6}Time and Frequency Geodesy Center (TFGC), School of Geodesy and Geomatics, Wuhan University, Wuhan 430079, China
}
\end{center}
\clearpage

\section{Per-segment data}

The campaign produced \nSegAll~jump-free segments in total, of which
\nSeg~enter the ratio combination. The excluded segment was recorded
while the experimental parameters were still being tuned: several of
its operating parameters and its Sr systematic shift take values far
from the other segments, and its ratio deviation is correspondingly
anomalous, so it is treated as preparation and testing rather than
as a measurement. The \nSegAll-segment combination including it
shifts the center value by $\dSegNineWls$ with a
comparable uncertainty; the exclusion does not drive the
reported value, but it removes a segment that is not representative
of the measurement conditions of the remaining
\nSeg~segments.

\begin{table}[h]
  \centering
  \caption{Per-segment data of the 16 segments analyzed in this work.
  Ratio deviations $y_i = R_i/R_1 - 1$ are relative to the first
  segment, with $R_1 = \RsegOne$ the segment-1 ratio used as the
  $y_i$ baseline. $u(y_i)$ is the statistical
  uncertainty of the segment ratio;
  ``Fit'' is \emph{est.} for segments whose stability fit lies in the
  white-frequency-noise band and \emph{lim.} for model-limited fits.}
  \label{tab:segments}
\begin{tabular}{rrrrl}
\toprule
Segment & $n_{\rm valid}$ & $y_i$ ($10^{-18}$) & $u(y_i)$ ($10^{-18}$) & Fit  \\
\midrule
1 & 68\,009 & +0.00 & 4.245 & est.  \\
2 & 29\,481 & -3.36 & 3.017 & lim.  \\
3 & 42\,891 & +2.00 & 5.880 & est.  \\
4 & 64\,280 & -4.51 & 3.674 & est.  \\
5 & 19\,532 & -3.18 & 15.250 & est.  \\
6 & 52\,140 & -1.05 & 2.230 & lim.  \\
7 & 39\,526 & +5.51 & 3.476 & lim.  \\
8 & 43\,565 & +1.07 & 9.860 & est.  \\
9 & 57\,938 & -0.76 & 5.003 & est.  \\
10 & 35\,824 & +2.61 & 5.976 & est.  \\
11 & 40\,346 & +0.63 & 4.843 & lim.  \\
12 & 152\,994 & -3.09 & 4.019 & est.  \\
13 & 68\,618 & +1.97 & 6.254 & est.  \\
14 & 57\,649 & -0.25 & 4.028 & est.  \\
15 & 147\,640 & -0.87 & 1.809 & lim.  \\
16 & 65\,983 & -1.09 & 8.813 & est.  \\
\bottomrule
\end{tabular}

\end{table}

\begin{table}[h]
  \centering
  \caption{Intermediate quantities of the per-segment ratio
  inversion: the retained sample count $n_{\rm valid}$, the
  segment-mean beat deviation $\overline{\delta b}$ and the Sr
  systematic shift $a$ entering the Dr formula, and the resulting
  ratio $R_i$ (24 digits shown; full precision in the frozen result
  layer).}
  \label{tab:ratio-intermediates}
\begin{tabular}{rrrrr}
\toprule
Segment & $n_{\rm valid}$ & $\overline{\delta b}$ (Hz) & $a$ & $R_i$  \\
\midrule
1 & 68\,009 & -0.025398 & -1.720e-16 & 1.2075070393433377166041  \\
2 & 29\,481 & -0.024746 & -1.720e-16 & 1.2075070393433377125522  \\
3 & 42\,891 & -0.025789 & -1.720e-16 & 1.2075070393433377190155  \\
4 & 64\,280 & -0.024502 & -1.720e-16 & 1.2075070393433377111564  \\
5 & 19\,532 & -0.024747 & -1.720e-16 & 1.2075070393433377127682  \\
6 & 52\,140 & -0.025168 & -1.720e-16 & 1.2075070393433377153358  \\
7 & 39\,526 & -0.026376 & -1.718e-16 & 1.2075070393433377232616  \\
8 & 43\,565 & -0.025530 & -1.717e-16 & 1.2075070393433377178917  \\
9 & 57\,938 & +0.009546 & +7.906e-18 & 1.2075070393433377156814  \\
10 & 35\,824 & +0.008892 & +7.906e-18 & 1.2075070393433377197594  \\
11 & 40\,346 & +0.009259 & +7.917e-18 & 1.2075070393433377173638  \\
12 & 152\,994 & +0.009998 & +7.917e-18 & 1.2075070393433377128695  \\
13 & 68\,618 & +0.009019 & +7.917e-18 & 1.2075070393433377189832  \\
14 & 57\,649 & +0.009448 & +7.917e-18 & 1.2075070393433377163028  \\
15 & 147\,640 & +0.009549 & +7.917e-18 & 1.2075070393433377155518  \\
16 & 65\,983 & +0.009573 & +7.917e-18 & 1.2075070393433377152824  \\
\bottomrule
\end{tabular}

\end{table}
\begin{table}[h]
  \centering
  \caption{Per-segment systematic shifts of the 16 segments analyzed
  in this work, as recorded in the campaign parameter file: the Sr
  components ($a_{\rm rou}$, $a_{\rm AC}$, $a_{\rm SM}$,
  $a_{\rm air}$, $a_{\rm BBR}$) with
  $a = a_{\rm rou} + a_{\rm AC} + a_{\rm SM} + a_{\rm air} +
  a_{\rm BBR}$, and the Yb shift $b_{\rm Yb}$; both enter the ratio
  inversion.}
  \label{tab:shift-components}
\begin{tabular}{rllllll}
\toprule
Segment & $a_{\rm rou}$ & $a_{\rm AC}$ & $a_{\rm SM}$ & $a_{\rm air}$ & $a_{\rm BBR}$ & $b_{\rm Yb}$  \\
\midrule
1 & $-2.341\times10^{-19}$ & $7.444\times10^{-18}$ & $-1.785\times10^{-16}$ & $-6.7\times10^{-19}$ & $0$ & $1.9\times10^{-18}$  \\
2 & $-1.476\times10^{-19}$ & $7.444\times10^{-18}$ & $-1.786\times10^{-16}$ & $-6.7\times10^{-19}$ & $0$ & $1.9\times10^{-18}$  \\
3 & $-2.447\times10^{-19}$ & $7.444\times10^{-18}$ & $-1.786\times10^{-16}$ & $-6.7\times10^{-19}$ & $0$ & $1.9\times10^{-18}$  \\
4 & $-2.819\times10^{-19}$ & $7.444\times10^{-18}$ & $-1.785\times10^{-16}$ & $-6.7\times10^{-19}$ & $0$ & $1.8\times10^{-18}$  \\
5 & $-2.722\times10^{-19}$ & $7.444\times10^{-18}$ & $-1.785\times10^{-16}$ & $-6.7\times10^{-19}$ & $0$ & $1.7\times10^{-18}$  \\
6 & $-2.665\times10^{-19}$ & $7.444\times10^{-18}$ & $-1.785\times10^{-16}$ & $-6.7\times10^{-19}$ & $0$ & $1.8\times10^{-18}$  \\
7 & $-2.53\times10^{-19}$ & $7.444\times10^{-18}$ & $-1.783\times10^{-16}$ & $-6.7\times10^{-19}$ & $0$ & $1.7\times10^{-18}$  \\
8 & $-2.335\times10^{-19}$ & $7.444\times10^{-18}$ & $-1.783\times10^{-16}$ & $-6.7\times10^{-19}$ & $0$ & $1.8\times10^{-18}$  \\
9 & $-3.532\times10^{-19}$ & $8.929\times10^{-18}$ & $0$ & $-6.7\times10^{-19}$ & $0$ & $1.9\times10^{-18}$  \\
10 & $-3.532\times10^{-19}$ & $8.929\times10^{-18}$ & $0$ & $-6.7\times10^{-19}$ & $0$ & $1.9\times10^{-18}$  \\
11 & $-3.422\times10^{-19}$ & $8.929\times10^{-18}$ & $0$ & $-6.7\times10^{-19}$ & $0$ & $2\times10^{-18}$  \\
12 & $-3.422\times10^{-19}$ & $8.929\times10^{-18}$ & $0$ & $-6.7\times10^{-19}$ & $0$ & $1.9\times10^{-18}$  \\
13 & $-3.422\times10^{-19}$ & $8.929\times10^{-18}$ & $0$ & $-6.7\times10^{-19}$ & $0$ & $1.9\times10^{-18}$  \\
14 & $-3.422\times10^{-19}$ & $8.929\times10^{-18}$ & $0$ & $-6.7\times10^{-19}$ & $0$ & $1.9\times10^{-18}$  \\
15 & $-3.422\times10^{-19}$ & $8.929\times10^{-18}$ & $0$ & $-6.7\times10^{-19}$ & $0$ & $2\times10^{-18}$  \\
16 & $-3.422\times10^{-19}$ & $8.929\times10^{-18}$ & $0$ & $-6.7\times10^{-19}$ & $0$ & $2.1\times10^{-18}$  \\
\bottomrule
\end{tabular}

\end{table}

\begin{table}[h]
  \centering
  \caption{Leave-one-out sensitivity of the ratio combination. Each
  row excludes one of the 16 segments and recomputes the combination
  from the remaining segments. The center values are stable at the
  $10^{-18}$ level under any single exclusion.}
  \label{tab:loo}
\begin{tabular}{rSSSS}
\toprule
Excluded segment & {$R_{\rm wls}$} & {$\chi^2_{\rm red}$} & {$u_{\rm wls}$ ($10^{-18}$)} & {$R_{\rm bayes}$}  \\
\midrule
1 & 1.2075070393433377161 & 0.554 & 1.136 & 1.2075070393433377161  \\
2 & 1.2075070393433377164 & 0.486 & 1.164 & 1.2075070393433377165  \\
3 & 1.2075070393433377160 & 0.543 & 1.123 & 1.2075070393433377160  \\
4 & 1.2075070393433377164 & 0.467 & 1.145 & 1.2075070393433377164  \\
5 & 1.2075070393433377161 & 0.553 & 1.111 & 1.2075070393433377161  \\
6 & 1.2075070393433377162 & 0.548 & 1.217 & 1.2075070393433377163  \\
7 & 1.2075070393433377155 & 0.345 & 1.150 & 1.2075070393433377156  \\
8 & 1.2075070393433377161 & 0.553 & 1.114 & 1.2075070393433377161  \\
9 & 1.2075070393433377161 & 0.554 & 1.128 & 1.2075070393433377161  \\
10 & 1.2075070393433377160 & 0.537 & 1.122 & 1.2075070393433377160  \\
11 & 1.2075070393433377160 & 0.551 & 1.130 & 1.2075070393433377160  \\
12 & 1.2075070393433377163 & 0.524 & 1.139 & 1.2075070393433377163  \\
13 & 1.2075070393433377160 & 0.545 & 1.121 & 1.2075070393433377160  \\
14 & 1.2075070393433377161 & 0.555 & 1.139 & 1.2075070393433377161  \\
15 & 1.2075070393433377163 & 0.549 & 1.287 & 1.2075070393433377163  \\
16 & 1.2075070393433377161 & 0.554 & 1.115 & 1.2075070393433377161  \\
\bottomrule
\end{tabular}

\end{table}

\begin{table}[h]
  \centering
  \caption{Per-segment selection diagnostics. Counts at each stage of
  the data processing chain: the observation window, the excluded
  intervals, the jump-filtered record, the endpoint-trimmed samples
  and the retained count.}
  \label{tab:selection}
\begin{tabular}{rrrrrr}
\toprule
Segment & window & excluded & jump-filtered & trimmed & retained  \\
\midrule
1 & 68\,011 & 0 & 68\,011 & 2 & 68\,009  \\
2 & 29\,491 & 1 & 29\,488 & 4 & 29\,481  \\
3 & 42\,899 & 0 & 42\,896 & 3 & 42\,891  \\
4 & 64\,286 & 1 & 64\,283 & 0 & 64\,280  \\
5 & 19\,538 & 0 & 19\,538 & 6 & 19\,532  \\
6 & 52\,154 & 0 & 52\,154 & 14 & 52\,140  \\
7 & 39\,600 & 1 & 39\,599 & 73 & 39\,526  \\
8 & 43\,607 & 0 & 43\,606 & 37 & 43\,565  \\
9 & 57\,959 & 0 & 57\,957 & 16 & 57\,938  \\
10 & 35\,841 & 0 & 35\,839 & 12 & 35\,824  \\
11 & 40\,374 & 0 & 40\,374 & 28 & 40\,346  \\
12 & 153\,000 & 0 & 153\,000 & 6 & 152\,994  \\
13 & 68\,624 & 0 & 68\,621 & 0 & 68\,618  \\
14 & 57\,653 & 0 & 57\,653 & 4 & 57\,649  \\
15 & 147\,650 & 0 & 147\,650 & 10 & 147\,640  \\
16 & 66\,013 & 0 & 66\,013 & 30 & 65\,983  \\
\bottomrule
\end{tabular}

\end{table}

\begin{figure}[p]
  \centering
  \includegraphics[width=\linewidth,height=0.85\textheight,keepaspectratio]{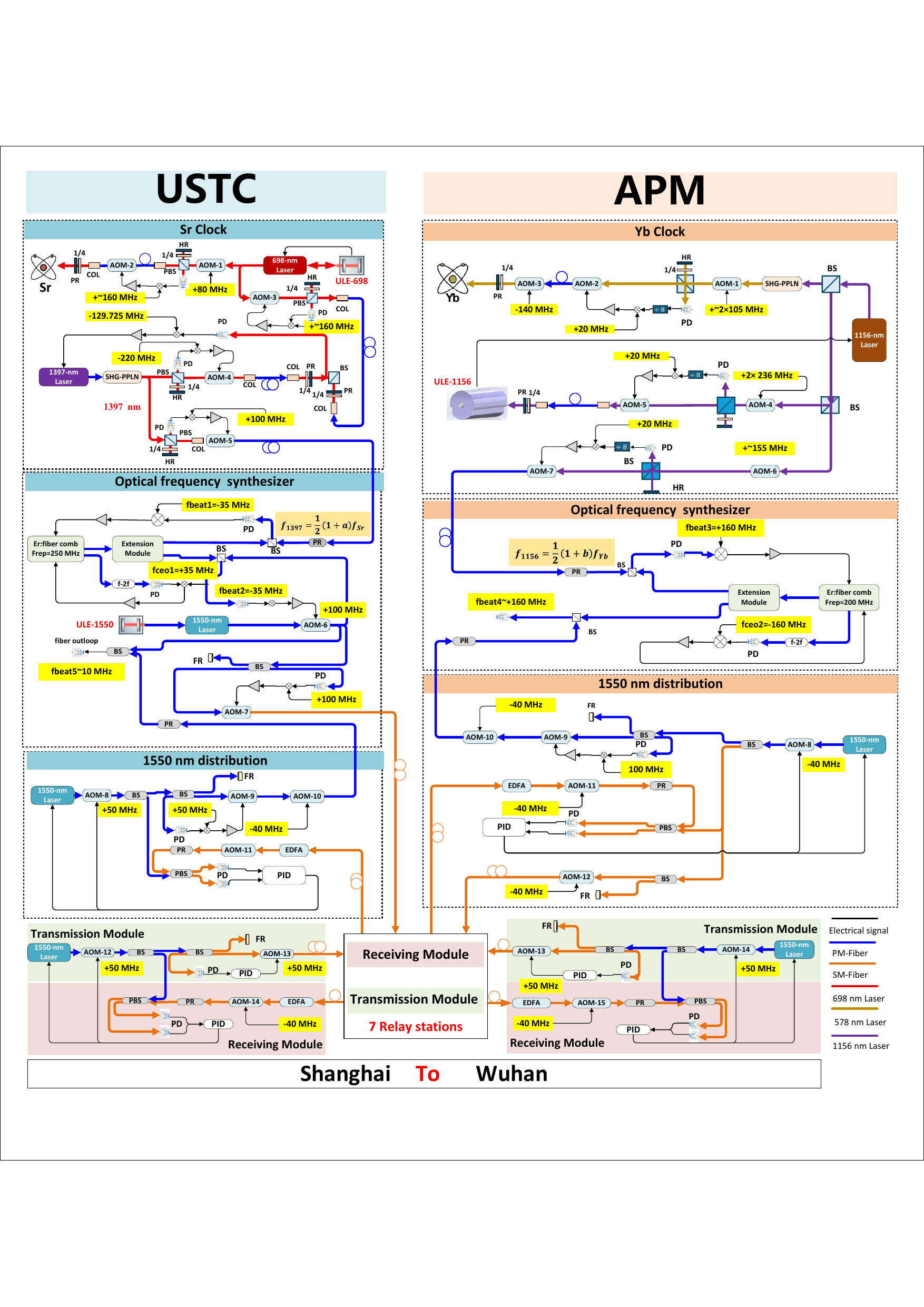}
  \caption{Detailed experimental scheme of the two-clock comparison.
  At each end the clock laser is stabilized to a high-finesse optical
  reference cavity and probes the respective clock transition; the
  clock light is then referenced to the local optical frequency comb,
  which bridges the clock wavelength to the \qty{1550}{nm} telecom
  carrier. Shanghai (left) locks the \qty{698}{nm} Sr clock laser,
  doubled to \qty{1397}{nm}, to the comb
  ($f_{\rm rep1}=\qty{250}{MHz}$), and a \qty{1550}{nm}
  ultra-stable laser to the same comb for transmission. Wuhan
  (right) locks the \qty{1156}{nm} Yb clock laser, doubled to
  \qty{578}{nm}, to its comb ($f_{\rm rep2}=\qty{200}{MHz}$), and
  beats the received carrier against it. AOM, acousto-optic
  modulator; ULE, ultra-low-expansion glass optical cavity; PBS,
  polarizing beam splitter; BS, beam splitter; PR, polarization
  rotator; HR, high-reflectivity mirror; COL, collimator; PD,
  photodetector; SHG-PPLN, second-harmonic generation in a
  periodically poled lithium-niobate crystal; EDFA, erbium-doped
  fiber amplifier; $f$--$2f$, carrier-envelope-offset detection.}
  \label{fig:method-setup}
\end{figure}

\section{Frequency-chain relations and the ratio inversion}
\label{sec:freq-chain}

The two clock transitions are bridged to the \qty{1550}{nm} telecom
carrier through fixed factor-of-two relations. On the Shanghai side
the Sr \qty{698}{nm} transition is doubled to \qty{1397}{nm}; on the
Wuhan side the Yb \qty{578}{nm} transition is doubled to
\qty{1156}{nm}. Both ends are phase-locked to independent optical
frequency combs. The Shanghai comb has a repetition rate of
\qty{250}{MHz}, derived from a \qty{10}{MHz} reference divided by
25, and the Wuhan comb \qty{200}{MHz}, divided by 20, as in the
Methods frequency relations.

The recorded observable is the beat $b$ between the
received carrier and the local Yb-referenced comb, sampled every
second. The ratio inversion uses the fixed chain constants
\begin{center}
\begin{tabular}{llll}
\toprule
Constant & Value & Meaning  \\
\midrule
$N_{1156}$ & 1,295,739 & Yb 1156 nm comb count  \\
$N_{1397}$ & 858,456 & Sr 1397 nm comb count  \\
$N_{1550}$ & 773,598 & 1550 nm link count (Shanghai end)  \\
$N_{1550}^{\rm WH}$ & 966,996 & 1550 nm link count (Wuhan end)  \\
$f_{\mathrm{ref}}$ & \qty{10}{MHz} & comb repetition base  \\
$\mathrm{div}_{20}$ & 20 & divider (Wuhan comb, $f_{\mathrm{rep}} = 200$ MHz)  \\
\bottomrule
\end{tabular}
\end{center}

The Shanghai 25-divider enters the inversion through the $7/25$ and
$1/25$ terms of the Dr formula below.

The beat is converted to fractional units with the transfer frequency
$f_{1550} = \fFifteenTHz$\,THz ($= N_{1550}^{\rm WH}\,f_{\mathrm{rep}}$).
The segment-mean beat deviation $\overline{\delta b}$ enters the
Sr/Yb ratio through
\[
\mathrm{Dr} = \frac{c_{1156}}{N_{1156}}\,
\frac{\overline{\delta b}}{f_{\mathrm{ref}}\,\mathrm{div}_{20}}
\bigg/ \left[
\frac{c_{1397}}{N_{1397}}
\left(N_{1550} + \frac{7}{25} + \frac{1}{25}\right)
\right],
\]
where $c_{1156} = (1+b_{\rm Yb})/2$ with $b_{\rm Yb}$ the Yb-side
systematic shift, applied per segment (see below)
and $c_{1397} = (1+a)/2$, with $a$ collecting the
five Sr systematic components listed below. The segment ratio is
\[
\mathrm{Sr/Yb} = \mathrm{ratio}_{\rm base} + \mathrm{Dr},
\qquad
\mathrm{ratio}_{\rm base} =
\frac{c_{1156}}{N_{1156}}\,
\frac{N_{1550}^{\rm WH} + \frac{26}{20} + \frac{m}{f_{\mathrm{ref}}\,\mathrm{div}_{20}}}
{\frac{c_{1397}}{N_{1397}}\left(N_{1550} + \frac{8}{25}\right)},
\]
where $m$ is the campaign-global median of the plausible beat samples.
The static gravitational-redshift correction is applied as a
multiplicative factor,
\[
(\mathrm{Yb/Sr})_{\rm corr} = (\mathrm{Yb/Sr})_{\rm raw}\,(1+\delta_g),
\qquad \delta_g = \deltaGV .
\]
All ratio arithmetic is carried out in 80-digit decimal precision,
mirroring the campaign processing convention.

The Sr systematic shift $a$ is the sum of five components evaluated
per segment: lattice AC-Stark ($a_{\rm rou}$, about
$-2.3$ to $-3.5\times10^{-19}$), AC Stark ($a_{\rm AC}$, $+7.44$ or
$+8.93\times10^{-18}$ by campaign period), second-order Zeeman
($a_{\rm SM}$, the dominant term), air refractive index
($a_{\rm air} = -6.7\times10^{-19}$), and blackbody radiation
($a_{\rm BBR} = 0$). The full per-segment values are listed in
Table~\ref{tab:shift-components}; the fixed chain constants are given
above, and the Yb-side systematic shift $b_{\rm Yb}$ is applied per
segment (16 recorded values, Table~\ref{tab:shift-components})
with the static gravitational correction
$\delta_g = \deltaGV$.

\section{Statistical model and combination}

Each segment $i$ contributes a ratio estimate $R_i$ and a statistical
uncertainty $u_i$ obtained from the overlapping Allan deviation
$\sigma_y(\tau)$ of the beat fractional frequency, evaluated on a tau
grid of powers of two plus 600, 1200, 3600 and 7200\,s, limited to a
quarter of the longest gap-free run of the segment. $u_i$ adopts the
last computable point, $u_i = R_i\,\sigma_y(\tau_{\max})$, without
extrapolating the stability curve beyond the data; the log-log slope
of $\sigma_y(\tau)$ is retained as a white-frequency-noise diagnostic
only.

The choice of the last computable point is deliberate and
conservative. The log-log instability slope flattens for
$\tau \gtrsim 4096$\,s (a flicker-frequency-noise floor from the
clocks and the link, not the tides); fitting the white-frequency
slope and extrapolating it to the full segment duration would
systematically understate $u_i$ by 30--40\%, inflating the reduced
chi-square to about 4 and making the segments appear mutually
inconsistent. Adopting $\sigma_y$ at the largest computable $\tau$
instead keeps the statistical uncertainty honest at the expense of
being somewhat conservative: the resulting $\chi^2_{\rm red} = 0.41$
and Birge ratio 0.64 are below unity, indicating that the segments
are consistent with their (slightly conservative) uncertainties.
Segments whose log-log slope leaves the white-frequency-noise band
$[-0.65, -0.35]$ are flagged as model-limited but retained with their
measured uncertainties; they lie at the steep end of a continuous
slope distribution with no common campaign, weather or laboratory
condition.

The segments are combined under a random-effect model with common
excess scatter, $R_i \sim \mathcal{N}(\mu, u_i^2 + \xi^2)$. Three
estimators are evaluated:

\begin{itemize}
\item \emph{Weighted least squares} with inverse-variance weights
$w_i = 1/u_i^2$: $\hat\mu = \sum w_i R_i/\sum w_i$, with
$u_{\rm wls} = 1/\sqrt{\sum w_i}$. The reduced chi-square
$\chi^2_{\rm red} = \sum (R_i-\hat\mu)^2/u_i^2/(N-1)$ and the Birge
ratio $\sqrt{\chi^2_{\rm red}}$ quantify the segment agreement.

\item \emph{Mandel--Paule}: the same observation model with
$\xi$ chosen so that $\chi^2_{\rm red}(\xi) = 1$.

\item \emph{Bayesian random effect}: flat priors on $\mu$ and on
$\log\xi$, with the joint posterior marginalized numerically over a
40-decade $\log\xi$ grid (401 points, trapezoidal quadrature). The
grid span is expanded until the posterior mass at the edges falls
below 1\%. The posterior mean of $\mu$ and its posterior standard
deviation give the reported central value and statistical
uncertainty.
\end{itemize}

All three center values are reconstructed in decimal precision from
the segment-1 baseline, $R = R_1\,(1+\hat\mu/R_1)$, so that
differences at the $10^{-18}$ level are not lost in floating-point
arithmetic.

\section{Tidal template construction}

The tidal perturbation to each clock is evaluated as an equivalent
gravitational-potential difference. For a station at geocentric
position $\mathbf{r}$, the degree-$n$ lunisolar potential is
\begin{equation*}
  W_n(\mathbf{r},t) = \sum_{b\in\{\mathrm{Sun},\mathrm{Moon}\}}
  \frac{G M_b}{r_b(t)}
  \left[\frac{r}{r_b(t)}\right]^{n} P_n\!\left(\cos\psi_b(t)\right),
\end{equation*}
where $r_b$ and $\psi_b$ are the geocentric distance of body $b$ and
the angle between $\mathbf{r}$ and $\mathbf{r}_b$. Sun and Moon
positions are taken from the JPL DE421 ephemeris, and all vectors
are referred to the ITRS through the full IERS 2010 transformation
(IAU 2006 precession, IAU 2000A nutation, Earth rotation angle and
polar motion), with $\mathrm{TT}-\mathrm{UTC}=69.184$\,s. The
degree-2 term dominates the clock signal; the degree-3 term is at
the millimeter level and is omitted.

The elastic response of the solid Earth is described by the
degree-2 Love numbers $h_2=0.6078$ and $k_2=0.3010$. For a clock
fixed to the deformed surface, the direct potential, the
mass-redistribution term and the geometric uplift combine to
\begin{equation*}
  \Delta W_{\mathrm{solid}}(t) = (1+k_2-h_2)
  \left[W_2^{\mathrm{M}}(t)+W_2^{\mathrm{S}}(t)\right]
  = 0.6932\left[W_2^{\mathrm{M}}(t)+W_2^{\mathrm{S}}(t)\right].
\end{equation*}

Ocean-tide loading is taken from the FES2022 loading-tide atlas
(AVISO), which provides the elastic vertical ground displacement
$\delta h_{\mathrm{otl}}$ on a $1/30^\circ$ global grid. The eight
major constituents (M$_2$, S$_2$, N$_2$, K$_2$, K$_1$, O$_1$,
P$_1$, Q$_1$) carry more than 99\% of the displacement power; their
amplitude and Greenwich phase lag are interpolated at each station
and evaluated as a harmonic series. The equivalent potential is
\begin{equation*}
  \Delta W_{\mathrm{otl}}(t) = -g\,\delta h_{\mathrm{otl}}(t),
  \qquad g = 9.798~\mathrm{m\,s^{-2}}.
\end{equation*}
The atlas already includes the load Love numbers, so no additional
load coefficient is applied. An independent evaluation with the
GOT4.10c loading-tide model agrees with FES2022 to 15--25\% in rms
displacement.

The comparison measures the differential potential between
Shanghai (\ang{31;14;} N, \ang{121;29;} E, $5$\,m) and Wuhan
(\ang{30;31;} N, \ang{114;19;} E, $30$\,m):
\begin{equation*}
  \Delta W(t) =
  \left[\Delta W_{\mathrm{solid}}+\Delta W_{\mathrm{otl}}\right]_{\mathrm{WH}}
  -
  \left[\Delta W_{\mathrm{solid}}+\Delta W_{\mathrm{otl}}\right]_{\mathrm{SH}}.
\end{equation*}
The corresponding fractional-frequency template is
$y_{\mathrm{tide}}(t)=\Delta W(t)/c^2$, or
$h(t)=f_{1550}\,y_{\mathrm{tide}}(t)$ in beat-frequency units, with
$f_{1550}=193.3992$\,THz. The template is computed on a
\qty{90}{s} Beijing-time grid and interpolated to the 1\,s beat
record. Over 20 June--10 September 2026, the solid-tide
differential has an rms of $2.85\times10^{-18}$ (peak-to-peak
$1.43\times10^{-17}$), the ocean-loading differential
$8.40\times10^{-19}$ ($4.31\times10^{-18}$), and their sum
$3.43\times10^{-18}$ ($1.66\times10^{-17}$). The static tide is
treated consistently by using the zero-tide gravity field XGM2019e;
a tide-free field would introduce a $4.5\times10^{-10}$ offset in
$W/c^2$.

\section{Within-segment tidal correlation analysis}

This section collects the per-segment results of the within-segment
response analysis. No single segment reaches a significant
correlation on its own: the per-segment fits scatter around zero
with uncertainties comparable to the template variation. Across the
\nSeg~segments entering the ratio combination, however, the sign of
the correlation is positive in \tidalNegSeg~of the \nSeg~segments,
and the precision-weighted response amplitude is $\tidalAmpNewE$
(DE421+FES2022), differing from zero at $\tidalSigmaNewE\sigma$ and
consistent with the expected response of $+1$ at the $1\sigma$
level, supporting the bias correction applied to the central
value.

\begin{table}[h]
  \centering
  \caption{Per-segment 1200-s-windowed tidal-correlation fits
  (DE421+FES2022 template). $A$ and $u_A$ are the fitted amplitude and
  its uncertainty, $r$ the Pearson correlation, all from
  non-overlapping 1200-s windows; the sign convention has the
  expected response at $A=+1$. The combined, precision-weighted
  amplitude is $\tidalAmpNewE$
  ($\tidalAmpNewUnc$, $\tidalSigmaNewE\sigma$ from zero),
  with the sign of the correlation positive in \tidalNegSeg~of the
  \nSeg~segments.}
  \label{tab:correlation}
\begin{tabular}{rrrrr}
\toprule
Segment & hours & $A$ & $u_A$ & $r$ \\
\midrule
1 & 18.9 & +1.41 & 0.52 & +0.39 \\
2 & 8.2 & +0.60 & 0.67 & +0.47 \\
3 & 11.9 & +1.84 & 0.54 & +0.54 \\
4 & 17.9 & +0.71 & 0.53 & +0.12 \\
5 & 5.4 & +1.81 & 1.33 & +0.40 \\
6 & 14.5 & +0.02 & 0.60 & -0.25 \\
7 & 11.0 & +0.72 & 0.84 & -0.04 \\
8 & 12.1 & -0.74 & 1.45 & -0.02 \\
9 & 16.1 & +1.15 & 0.73 & +0.21 \\
10 & 10.0 & -0.03 & 1.96 & +0.03 \\
11 & 11.2 & +0.66 & 0.87 & +0.19 \\
12 & 42.5 & +0.46 & 0.39 & +0.03 \\
13 & 19.1 & +1.21 & 0.32 & +0.53 \\
14 & 16.0 & -0.15 & 1.16 & -0.07 \\
15 & 41.0 & +0.09 & 0.69 & +0.01 \\
16 & 18.3 & +1.09 & 0.63 & +0.28 \\
\bottomrule
\end{tabular}

\end{table}

\begin{figure}[p]
  \centering
  \includegraphics[width=\linewidth,height=0.8\textheight,keepaspectratio]{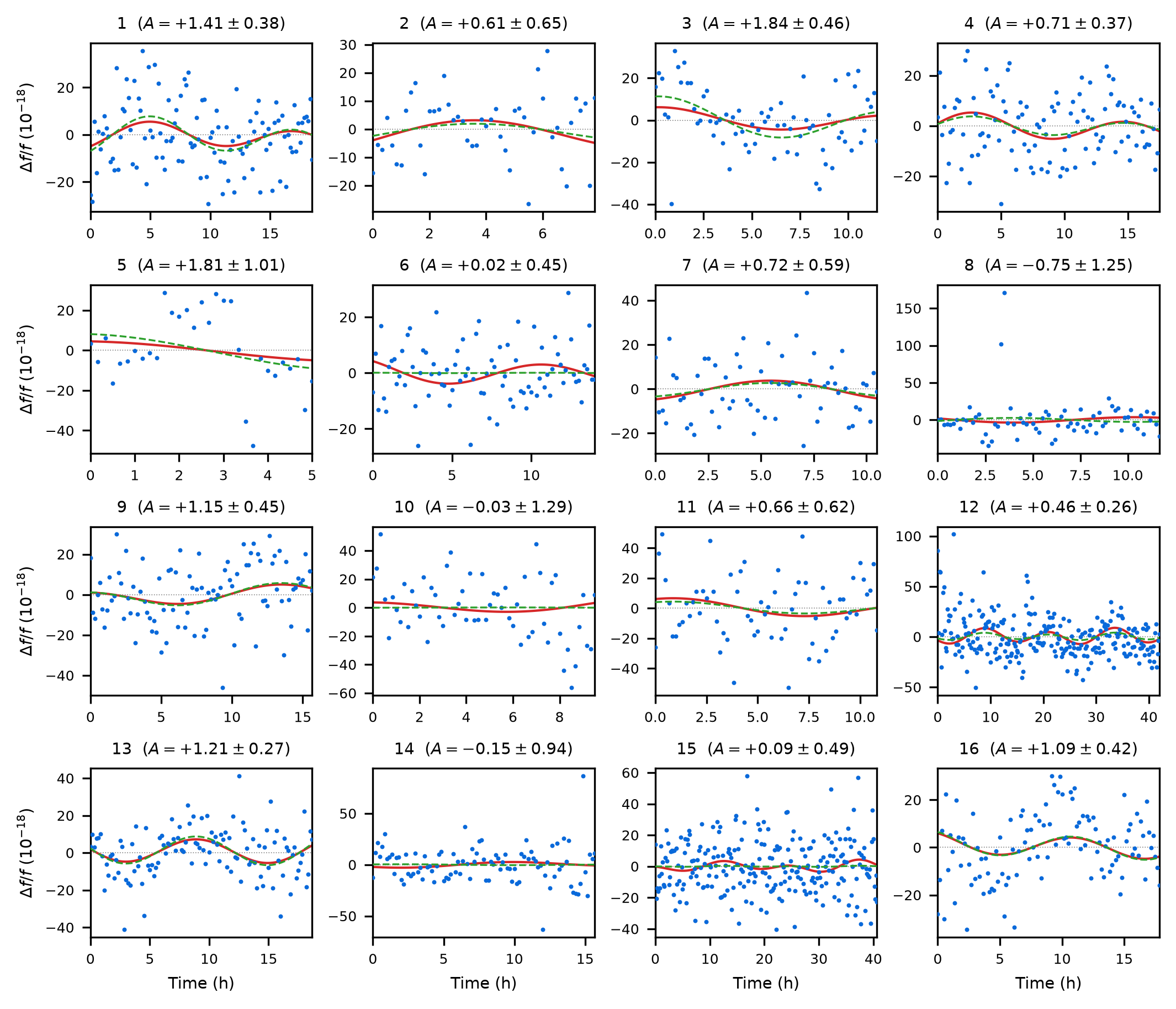}
  \caption{Per-segment fits of the 1200-s-windowed beat against the
  DE421+FES2022 template (sign convention with the expected response
  at $A=+1$). Each panel shows the measured beat, the template, and
  the fitted line with its amplitude; no single segment is
  individually significant. The combined amplitude is
  $\tidalAmpNewE$ (Methods Figure~5).}
  \label{fig:fit-grid}
\end{figure}

Environmental checks accompany this analysis. Co-seismic
displacement from the regional seismicity
during the campaign is estimated from the earthquake moments to be
at most $6\times10^{-20}$ in differential fractional frequency, two
orders below the statistical uncertainty, so it cannot drive the
tidal-period correlation.

\begin{table}[h]
  \centering
  \caption{Tidal combination summary (16 segments): the reduced
  chi-square and Birge ratio of the segment combination, the Bayesian
  excess scatter $\xi$, and $u_{\rm wls}$, for the uncorrected record
  and for the fixed
  theoretical correction amplitude ($A=+1$). The correction degrades
  the combination statistics.}
  \label{tab:tidal-synthesis}
\begin{tabular}{lrrrr}
\toprule
Scenario & $\chi^2_{\rm red}$ & Birge & $\xi_{\rm Bayes}$ ($10^{-18}$) & $u_{\rm wls}$ ($10^{-18}$) \\
\midrule
uncorrected ($A=0$) & 0.421 & 0.649 & 0.14 & 1.128 \\
theory ($A=+1$) & 0.602 & 0.776 & 0.14 & 0.862 \\
\bottomrule
\end{tabular}

\end{table}

\end{document}